\documentclass[12pt,lettersize,journal,onecolumn]{IEEEtran}
\usepackage{amsmath,amsfonts}
\usepackage{algorithmic}
\usepackage{algorithm}
\usepackage{array}
\usepackage[caption=false,font=normalsize,labelfont=sf,textfont=sf]{subfig}
\usepackage{textcomp}
\usepackage{stfloats}
\usepackage{url}
\usepackage{verbatim}
\usepackage{graphicx}
\usepackage{cite}
\usepackage{cases}
\usepackage{epstopdf}
\usepackage{float}
\usepackage{xcolor}
\usepackage{setspace}
\begin{document}

\title{Free-space Continuous-variable Quantum Secret Sharing}

\author{Fangli Yang, \and Liang Chang, \and Daowen Qiu, \and Minghua Pan, \and Wanjun Xiong
\thanks{Manuscript created October, 2024; This work was supported by the National Natural Science Foundation of China (Grant Nos. U22A2099, 62361021). \textit{(Corresponding author:
Liang Chang).}}
\thanks{F. Yang and L. Chang are with Guangxi Key Laboratory of Trusted Software, Guilin University of Electronic Technology, Guilin 541004, China (email: changl@guet.edu.cn).}
\thanks{D. Qiu is with Institute of Quantum Computing and Software, School of Computer Science and Engineering, Sun Yat-sen University, Guangzhou 510006, China.}
\thanks{M. Pan and W. Xiong are with Guangxi Key Laboratory of Cryptography and Information Security, Guilin University of Electronic Technology, Guilin 541004, China.}
}

\markboth{Journal of \LaTeX\ Class Files,~Vol.~, No.~, October~2024}%
{Shell \MakeLowercase{\textit{et al.}}: A Sample Article Using IEEEtran.cls for IEEE Journals}

\IEEEpubid{0000--0000/00\$00.00~\copyright~2021 IEEE}

\maketitle

\begin{abstract}
Free-space quantum cryptography has the potential to enable global quantum communication. However, most existing continuous-variable quantum secret sharing (CV-QSS) schemes rely on fiber channels. In this paper, we present a CV-QSS protocol designed for free-space transmission and construct models of crucial parameters, including channel transmittance, excess noise, and interruption probability, thus deriving the bound of the secret key rate. In particular, we provide a multi-source excess noise model for free-space CV-QSS based on the local local oscillator (LLO) scheme and a straightforward optimization of the noise.
Furthermore, our research considers the impact of atmospheric turbulence on the protocol. 
Simulation results demonstrate that the intensification of atmospheric turbulence adversely affects the aforementioned crucial parameters, leading to a significant reduction in the key rate.
However, our protocol shows its capability to securely share secrets over a distance of exceeding $60$ km among five participants in the presence of turbulence with $C^2_n=1\times 10^{-15}m^{-2/3}$, while maintaining a high key rate of approximately $0.55$ bit/pulse over a distance of $10$ km across twenty participants. These findings suggest that efficient CV-QSS in free space is indeed achievable. This research may serve as a reference for the design and optimization of the practical CV-QSS system.
\end{abstract}

\begin{IEEEkeywords}
Quantum secret sharing, Continuous-variable, Free-space channel, Atmospheric turbulence.
\end{IEEEkeywords}

\section{Introduction}
\IEEEPARstart{S}{ecret} sharing \cite{1979How,PhysRevLett.95.230505} plays an important role within the realm of information security. In a secret sharing system, a legitimate user, called as the dealer, divides a secret into $n$ sub-secrets (or subkeys) and distributes them to $n$ participants for safekeeping, requiring no less than $k\leq n$ participants to join forces to recover the secret. Secret sharing finds widespread application in E-Business, banking, and politics as a means of ensuring the confidentiality and integrity of sensitive communications.

As quantum information technology continues to advance based on quantum mechanics \cite{bennett2000quantum,wilde2013quantum}, Hillery, Buzek and Berthiaume introduced secret sharing into the quantum domain, presenting the first quantum secret sharing (QSS) scheme by using three-particle GHZ states, also known as the HBB protocol \cite{hillery1999quantum} in 1999. Since then, numerous achievements \cite{gottesman2000theory,PhysRevA.65.042310,markham2008graph,senthoor2022theory,ouyang2023approximate,conlon2024verifying,singh2024controlled} have been made in QSS. Presently, there are two forms of QSS implementation: discrete-variable (DV)-QSS based on a finite-dimensional Hilbert space and continuous-variable (CV)-QSS in an infinite-dimensional Hilbert space.
Compared to DV-QSS, which relies on single-photon sources that are challenging to prepare and detect, CV-QSS offers the advantage of being more practical and compatible with conventional communications. This is due to the fact that quantum signals can be effectively prepared, modulated, and measured in quantum optics using continuous variable quadratures.
Based on its superiorities, CV-QSS has been well developed in recent years. For example, \cite{lau2013quantum} proposed a CV-QSS protocol with continuous-variable cluster states, and \cite{kogias2017unconditional} provided an unconditional security proof of CV-QSS against both eavesdroppers and dishonest participants based on multiparty entanglement. In particular, \cite{grice2019quantum} showed an easy-to-implement CV-QSS utilizing weak coherent states and the balanced detection. Furthermore, \cite{grice2019quantum} applied the continuous-variable quantum key distribution (CV-QKD) \cite{Nature421,xu2020secure,yamano2024finite,hajomer2024long} technique to analyse the security and key rate of CV-QSS. Based on this work, scholars have proposed CV-QSS protocols from different perspectives. In \cite{PhysRevA.101.022301}, the authors studied the CV-QSS with thermal sources and discussed the finite-size effect. \cite{PhysRevA.103.032410,liao2023continuous} showed the CV-QSS protocols by using discrete modulated coherent states. \cite{arXiv2024Liao} described a practical CV-QSS scheme using LLO for two participants. However, these studies generally focus on the protocols within conventional fiber channels.

The practical implementation of CV-QSS may be restricted by birefringence effects and inherent losses in fiber channels. An alternative channel model for quantum cryptography is the free-space channel, which virtually eliminates the birefringence effect, preserves non-classical effects, and offers the possibility of broader geographical coverage. Recently, \cite{liu2021continuous} presented a CV-QSS protocol for a wireless link with a terahertz source frequency. \cite{yang2023continuous} investigated a CV-QSS protocol with a special feature: the channel transmittance varies according to a uniform probability distribution. However, none of the previous studies discussed the atmospheric effects of free-space channels on CV-QSS in detail. It is important to note that atmospheric channels can have negative effects on system performance due to turbulence \cite{vasylyev2016atmospheric,PhysRevA.99.053830,trinh2022statistical}.

Actually, atmospheric effects on CV-QKD have been extensively researched \cite{wang2018atmospheric,ruppert2019fading,zuo2020atmospheric,dequal2021feasibility,ghalaii2022quantum,10415457}. CV-QKD can distribute secret keys between only two remote parties over an insecure quantum channel. In a CV-QSS with $n$ participants, the dealer will establish $n$ independent QKD links with each participant to generate $n$ subkeys $\{K_1, K_2,\cdot\cdot\cdot, K_n\}$. It follows that CV-QSS can be seen as an application of CV-QKD. However, CV-QSS involves multiple participants and the modeling of its parameters is a more complex undertaking, necessitating a comprehensive examination of the structural characteristics of the entire protocol.

Based on the background provided above, we present an analysis of the feasibility of a free-space CV-QSS protocol with LLO and model the principal aspects of the protocol, namely transmittance, excess noise, and interruption probability.
In particular, the transmittance is primarily examined in terms of beam extinction and atmospheric turbulence. For this purpose, an elliptical model is used to derive an expression, and the Monte Carlo method is employed to estimate the expectation and variance.
In conjunction with the structural characteristics of CV-QSS, an evaluation of the various sources of excess noise is conducted, and an optimization is implemented with the objective of reducing noise. 
As for the non-interruption probability of the CV-QSS system, we consider the fact that all QKD links must not be interrupted so that the secret sharing process can be completed. Furthermore, this paper analyzes and contrasts three crucial aspects and the performance of the free-space CV-QSS system in varying turbulence intensities through numerical simulations. 

The rest of the paper is organized as follows. In Section \ref{sec:2}, the free-space CV-QSS with LLO is described in detail. In Section \ref{sec:3}, we study the channel transmittance, excess noise, and interrupt probability of CV-QSS in free-space channels. In Section \ref{sec:4}, we discuss the secret key rates of the free-space CV-QSS protocol by simulations and comparisons. The conclusion is given in Section \ref{sec:5}.

\section{Free-space CV-QSS system description}\label{sec:2}
It is widely recognized that the coherent detection of quantum signal pulses requires the use of high-power local oscillators (LO). In a continuous-variable (CV) system, the quantum signal and LO are often produced by a single laser at the sender and transmitted through a quantum channel, known as a trusted LO (TLO) system. This implementation leaves security loopholes that can be exploited by eavesdroppers to carry out attacks \cite{zhang2024continuous}. In this system, we utilize the local LO (LLO) \cite{hajomer2024long,qi2015generating}, which is generated by the dealer, thus circumventing the transmission of the LO through an insecure quantum channel.

\begin{figure}[!t]
\centering
\includegraphics[width=4in]{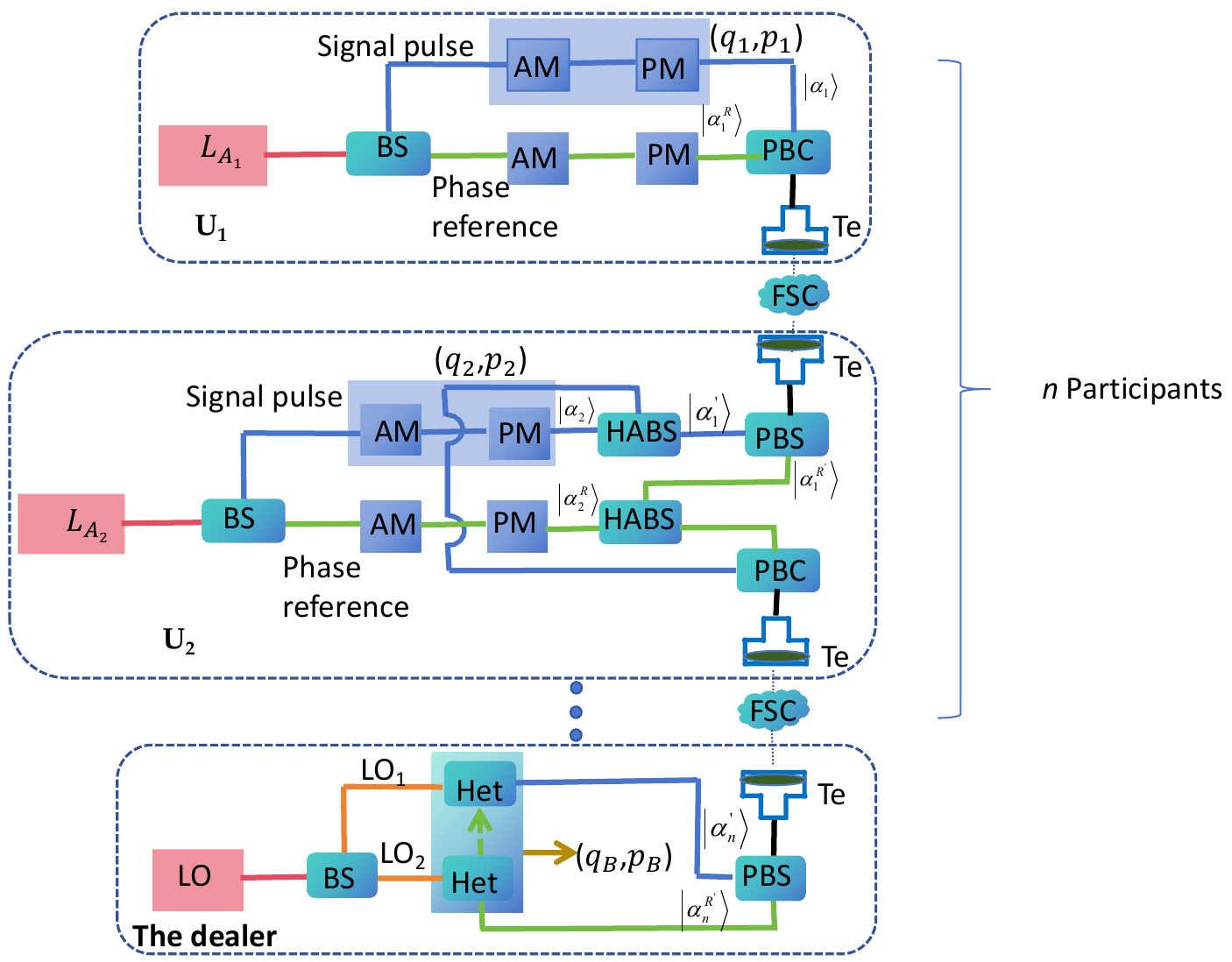}
\caption{The structure of the free-space CV-QSS with LLO, comprising a dealer and $n$ participants, denoted as $U_1, U_2,\cdot\cdot\cdot, U_n$. BS: beam splitter, AM: amplitude modulator, PM: phase modulator, PBC: polarized beam combine, PBS: polarized beam splitter, HABS: highly asymmetric beam splitter, Het: heterodyne detection.}
\label{cvqsslink}
\end{figure}

The structure of the free-space CV-QSS protocol with LLO is illustrated in Fig. \ref{cvqsslink}. It consists of a dealer and $n$ participants, designated as $U_1, U_2, \dots, U_n$. The protocol's procedure can be divided into two distinct stages: the quantum stage and the classical post-processing stage. It is important to note that, due to the free-space nature of the channel, telescopes (Te) are necessary for both signal transmission and reception. For brevity, this step is not described again in the description of CV-QSS since it is already implemented by default.
\subsection{Quantum stage}
At each $U_j$ $(j=1,\cdot\cdot\cdot,n)$, a laser light $L_{A_j}$ is used to generate the signal pulse and a phase reference pulse by a beam splitter (BS). The signal pulse is modulated with randomly Gaussian quadrature values $q_j$ and $p_j$ by amplitude modulator (AM) and phase modulator (PM), so as to form a Gaussian modulated coherent state $|\alpha_j\rangle$ centered on $\mathbf{x_j}=(q_j, p_j)$ in phase space. The phase reference pulse is encoded with predetermined values and it is denoted as $|\alpha^{R}_j\rangle$. Firstly, $U_1$ multiplies the pair $\{|\alpha_1\rangle,|\alpha^{R}_1\rangle\}$ of coherent states by a polarized beam combine (PBC) and transmits it to the next neighbor participant $U_2$ via a free-space channel (FSC). Secondly, $U_2$ de-multiplexes it to $|\alpha'_1\rangle$ and $|\alpha^{R'}_1\rangle$ by a polarized beam splitter (PBS), and couples his (or her) local state $|\alpha_2\rangle$ to the same spatiotemporal mode as $|\alpha'_1\rangle$ by using a highly asymmetric beam splitter (HABS) of transmissivity $T_H$, while the phase references $|\alpha^R_2\rangle$ and $|\alpha^{R'}_1\rangle$ are treated as such. Then, $U_2$ multiplies the superimposed signal and reference by a PBC and transmits it to the next participant. The process continues with the remaining participants. Subsequently, the dealer de-multiplexes it into the signal $|\alpha'_n\rangle$ and the phase reference $|\alpha^{R'}_n\rangle$, where the signal $|\alpha'_n\rangle$ is centered on
\begin{equation}
\begin{split}
\mathbf{x_B}&=\left(q_B, p_B\right)\\
&=\left(\sum_{j=1}^{n}\sqrt{T_j}q_{j},\sum_{j=1}^{n}\sqrt{T_j}p_{j}\right).
\end{split}
\end{equation}
$T_{j}$ is the channel transmittance between $U_j$ and the dealer, including the effects of HABS and free-space factors.
Besides, the dealer generates a light LO and splits it into $LO_1$ and $LO_2$ for coherent detection of the signal and the phase reference.
The coherent (heterodyne) measurement results $\{q_B, p_B\}$ of the signal are kept as raw data. Finally, repeat the above steps $N$ times to generate a set of raw data $A_N=\{\{q_{B_1},p_{B_1}\},\cdot\cdot\cdot,\{q_{B_i},p_{B_i}\},\cdot\cdot\cdot,\{q_{B_N},p_{B_N}\}\}$.

\subsection{Classical post-processing stage}
The dealer initially selects a random subset $B_n$ consisting of $n$ pairs from $A_N$ in order to estimate the $n$ channel transmittances $\{T_1,T_2,\cdot\cdot\cdot,T_n\}$.
Then, the dealer picks a pair $\{q_{B_i},p_{B_i}\}$ at random from the remaining raw data set $A_N/B_n$ and instructs all participants except for $U_j$, who is chosen as the honest one, to disclose their corresponding random numbers. 
By utilizing the announced data and $\{T_1,T_2,\cdot\cdot\cdot,T_n\}$, the dealer is able to calculate the pair $\{q'_{i_j},p'_{i_j}\}$. At this point, $U_j$ and the dealer can be analogized to Alice and Bob, respectively, serving as the trusted entities in a point-to-point quantum key distribution (QKD) link, denoted as $Link_j$.
By utilizing $\{q'_{i_j},p'_{i_j}\}$ and $U_j's$ data $\{q_{i_j},p_{i_j}\}$, we are able to derive a lower bound of secure key rate $r_{j}$. Subsequently, this iterative process is replicated $n$ times to establish a total of $n$ secure QKD links. In each iteration, a distinct participant is designated as the honest party.
After that, employing the standard CV-QKD protocol \cite{Nature421}, the dealer shares the corresponding security key $K_j$ with each participant $U_j$ against all the other $n-1$ participants and potential eavesdroppers in the channel from the remaining undisclosed data, where $j=1,\cdot\cdot\cdot,n$.
Finally, the dealer generates a new key $K = K_1\oplus K_2 \oplus\cdot\cdot\cdot\oplus K_n$ and encrypts the message M via $E = M\oplus K$. 

\section{Parameters modeling of free-space CV-QSS}\label{sec:3}
In this section, we model and discuss in depth the essential parameters of free-space CV-QSS, namely channel transmittance, channel excess noise, and communication interruption probability. 
\subsection{Channel transmittance}\label{sec:3.1}
A free-space channel can be decomposed into a set of independent subchannels, where the transmittance $t$ and the excess noise $\epsilon(t)$ of each subchannel can be regarded as stable, with the corresponding probability distribution $p(t)$ such that $\int p(t)dt=1$ \cite{hosseinidehaj2021composable}. For the entire free-space channel over all subchannels, we have $\langle t\rangle=\int tp(t)dt$ and $\langle \epsilon\rangle=\int \epsilon(t) p(t)dt$.

We use a vector $\textbf{T}=(T_1,\cdot\cdot\cdot,T_j,\cdot\cdot\cdot,T_n)$ to describe the characteristics of the free-space channel in the CV-QSS system. Here, $T_j$ represents the transmittance of $Link_j$ with a probability distribution $p(T_j)$, and any $T_i$ and $T_j$ are independent of each other ($i\neq j, 1\leq i,j\leq n$). Therefore, the probability density function of $\textbf{T}$ can be expressed as $p(\textbf{T})=p(T_1)\times\cdot\cdot\cdot \times p(T_n)=\prod \limits_{j=1}^np(T_j)$, and the expectation is derived as 
\begin{equation}
\begin{split}
\langle \textbf{T}\rangle&=\int_{\mathbb{R}^n} \textbf{T}p(\textbf{T})d\textbf{T}\\
&=\left(\langle T_1\rangle,\cdot\cdot\cdot,\langle T_j\rangle,\cdot\cdot\cdot,\langle T_n\rangle\right). 
\end{split}
\end{equation}

Free-space effects on transmittance are mainly due to atmospheric turbulence and beam extinction \cite{wang2018atmospheric}. In this paper, we consider a horizontally linked CV-QSS system at the Earth's surface. Assuming that the transmittance of atmospheric turbulence and beam extinction is $T_{at}$ and $T_{ex}$, respectively, then the transmittance of the $Link_j$ is 
\begin{equation}\label{Tall}
T_j=T_{at,j}T_{ex,j}T_H^{f(j,n)},
\end{equation}
where 
\begin{numcases}{f(j,n)=}
    n-j, & $j = 1 $  \\
    n-j+1, & $j = 2,\cdot\cdot\cdot,n. $
\end{numcases}

Absorption and scattering by molecules and aerosols lead to beam extinction, which causes the transmittance to decrease as the transmission distance increases, i.e., the transmittance associated with beam extinction is
\begin{equation}\label{extinction}
T_{ex,j}=e^{-\alpha_{\lambda}(h)L_j},
\end{equation}
where $L_j=L\times(n-j+1)/n$ denotes the horizontal distance between participant $U_j$ and the dealer. For ease of calculation, this paper assumes that all neighboring participants are equidistant, and the analysis method is similar if they are not equidistant, so it will not be repeated here. The equation $\alpha_{\lambda}(h)=N(h)\sum_{i=1}^{4}\alpha_{\lambda}^{i}$ describes the total extinction coefficient at a given height $h$ above sea level, where $\alpha_{\lambda}^{i}(i=1,\cdot\cdot\cdot,4)$ represents the extinction coefficient caused by aerosol scattering, aerosol absorption, molecular scattering, and molecular absorption, respectively. For the optical wavelength $\lambda=800$ nm, the total extinction coefficient can be estimated as $\alpha_{\lambda}(h)=\beta_0 e^{(-h/h_0)}$, where $h_0=6600$ m and $\beta_0=5\times 10^{-6}$ $\rm{m}^{-1}$ \cite{pirandola2021limits}.

\begin{figure}[!t]
\centering
\includegraphics[width=3in]{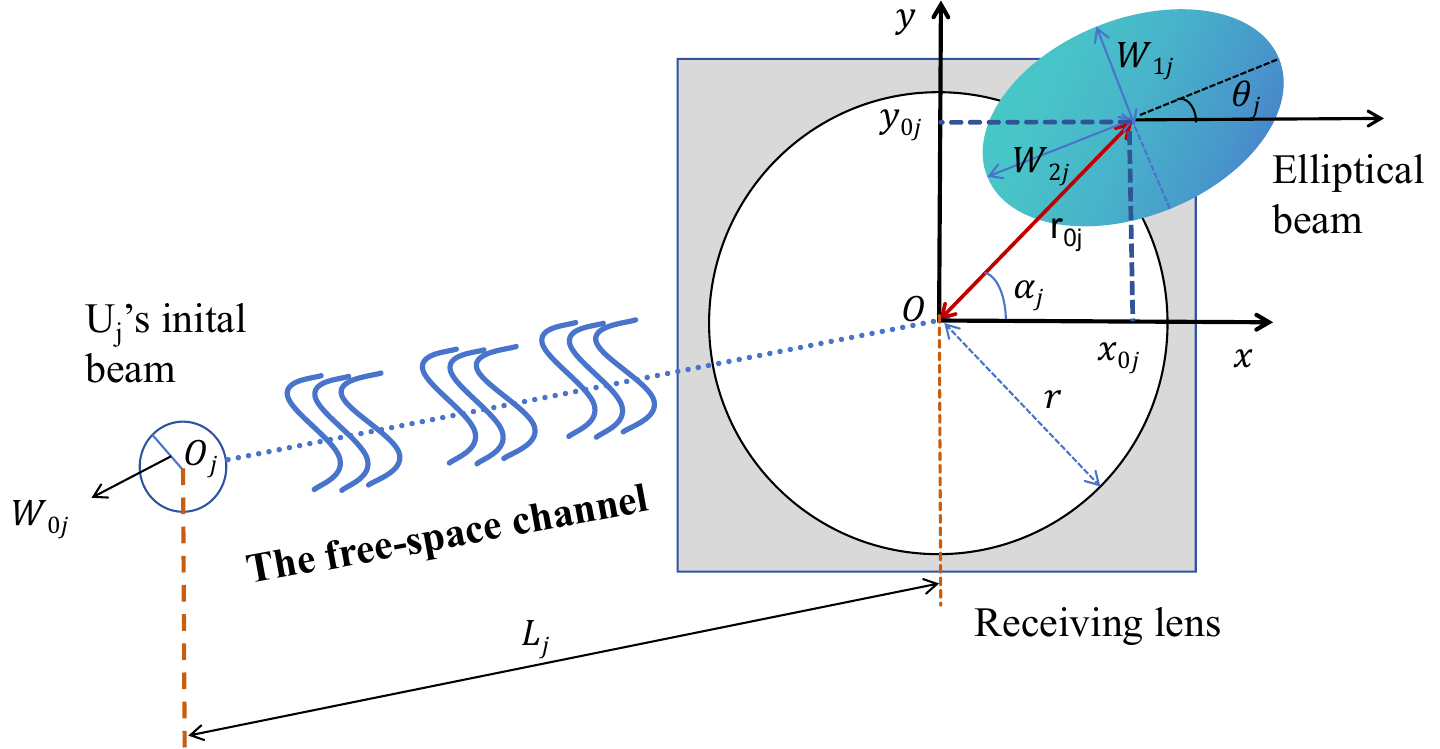}
\caption{The elliptical beam model. $U_j$'s inital Guassian beam is transformed into an elliptical shape after passing through a turbulence channel in free space. $O_j$: beam-centroid position of $U_j$'s Guassian beam, O: beam-centroid position of the dealer's telescope.}
\label{tuoyuan}
\end{figure}

In the atmospheric turbulence channel, the transmittance is randomly jittered due to beam wandering, broadening, deformation, and scintillation. The elliptical beam model \cite{vasylyev2016atmospheric}  describes the atmospheric turbulence well, with a transmittance probability distribution closer to the real experimental data than the negative logarithmic Weibull model. The security and feasibility of atmospheric quantum communication based on the elliptic model has been analyzed and proved in recent years \cite{wang2018atmospheric,zuo2020atmospheric,PhysRevA.99.032326}. Therefore, the elliptic model is used to describe the atmospheric channel in the CV-QSS.
As shown in Fig. \ref{tuoyuan}, the spot where the signal beam reaches the receiver is approximated as an ellipse in this model, and the channel transmittance probability distribution can be calculated by deriving the Glauber-Sudarshan P-function \cite{vasylyev2016atmospheric}.  

For $Link_j$, the elliptical spot at the receiving aperture plane can be described by a four-dimensional Gaussian random variable {\small $\textbf{v}=\{x_{0j},y_{0j},W_{1j},W_{2j}\}$} and an independent uniformly distributed variable $\theta_j$, where $(x_{0j},y_{0j})$ denotes the centroid position of the ellipse in the rectangular coordinate system with $O$ as the origin. {\small $W_{1j}=\sqrt{W_{0j}^2 {\rm exp}(\phi_{1j})}$} and {\small $W_{2j}=\sqrt{W_{0j}^2 {\rm exp}(\phi_{2j})}$} are semi-axes of the elliptical spot, where $W_{0j}$ is the $U_j$'s Gaussian beam-spot radius and $\phi_{ij}(i=1,2)$ are variables that conform to normal distributions. $\theta_j\in [0,\pi/2]$ is the counterclockwise angle between the long semi-axis and the $x$ axis.
In this model, the ellipse center position $(x_{0j},y_{0j})$ is related to beam wandering, while {\small $\{W_{1j},W_{2j},\theta_j\}$} can be used to describe beam broadening and deformation. The transmittance of $Link_j$ caused by the atmospheric turbulence can be modeled by \cite{vasylyev2016atmospheric}
\begin{equation}\label{Tj}
\small
T_{at,j}=T_{0j} {\rm exp}\left\{-\left[\frac{r_{0j}/r}{R\left(\frac{2}{{\rm W_{eff}}(\theta_j-\alpha_j)}\right)}\right]^{Q\left(\frac{2}{{\rm W_{eff}}(\theta_j-\alpha_j)}\right)}\right\},
\end{equation}
where $r$ is the receiving aperture radius and {\small $r_{0j}=\sqrt{x^2_{0j}+y^2_{0j}}$. $T_{0j}$} is the transmittance for the centered beam ($r_{0j}=0$) and {\small ${\rm W_{eff}}(\cdot)$} is the effective squared spot radius. Appendix A shows the derivation of $T_{0j}$ and {\small ${\rm W_{eff}}(\cdot)$}.

Eq. \ref{Tj} shows that $T_{at,j}$ is related to both a four-dimensional Gaussian random variable $\textbf{w}=\{x_{0j},y_{0j},\phi_{1j},\phi_{2j}\}$ and the uniform random variable $\theta_j$, where the covariance matrix elements of $\textbf{w}$ are described in Appendix \ref{elementsw}. Based on these random variables and Eq. \ref{Tj}, the expectations of $T_{at,1}$ and  $\sqrt{T_{at,1}}$ can be estimated by  Monte Carlo simulations. Further, according to Eq. \ref{Tall}, $\langle T_{j} \rangle$ and $\langle \sqrt{T_{j}} \rangle$ can be obtained. The other parameters are expressed in Table \ref{table1}. 

\begin{figure}[!t]
\centering
\includegraphics[width=3in]{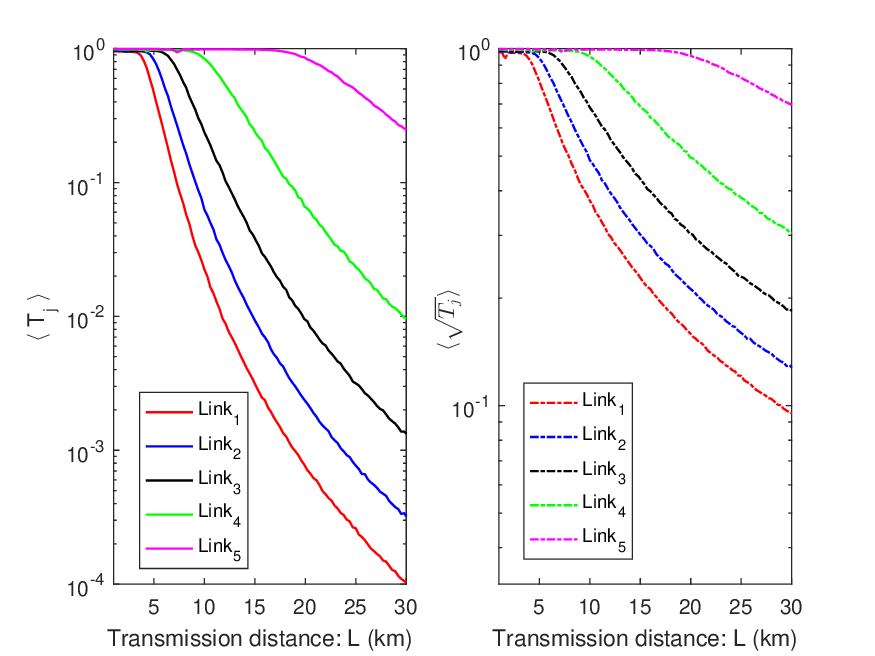}
\caption{$\langle T_j \rangle$ and $\langle \sqrt{T_j} \rangle$ as the fuctions of transmission distance L with $n=5$ and $C^2_n=5\times 10^{-15}m^{-2/3}$.}
\label{et}
\end{figure}

\begin{figure}[!t]
\centering
    \includegraphics[width=3in]{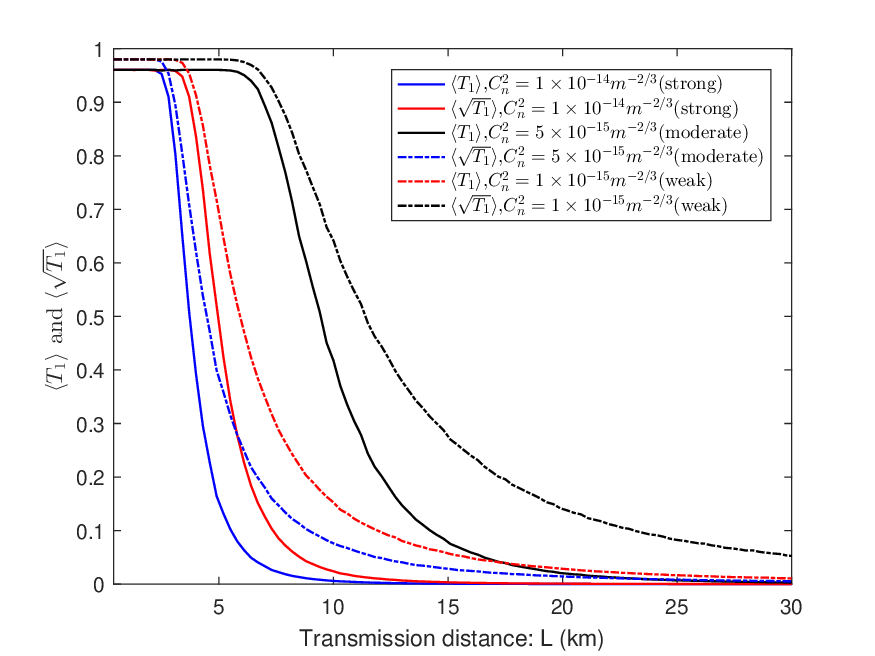}
    \caption{The statistical properties of transmittance $T_1$ in different turbulence intensities: 
$C^2_n=1\times 10^{-15}m^{-2/3}$(weak), $5\times 10^{-15}m^{-2/3}$(moderate), $1\times 10^{-14}m^{-2/3}$(strong).}
\label{sqrtT1andT1}
\end{figure}

Fig. \ref{et} illustrates the variations of $\langle T_{j} \rangle$ and $\langle \sqrt{T_{j}} \rangle$ with transmission distance $L$. The simulation results show that among the $n$ QKD links of CV-QSS, $Link_1$ has the smallest transmittance expectation, due to the fact that $U_1$ is farthest away from the dealer. We present $\langle T_{1} \rangle$ and $\langle \sqrt{T_{1}} \rangle$ in Fig. \ref{sqrtT1andT1}, when classify the values of the atmospheric turbulence intensity to $\{C^2_n=1\times 10^{-15}m^{-2/3}({\rm weak}), 5\times 10^{-15}m^{-2/3}({\rm moderate}), 1\times 10^{-14}m^{-2/3}({\rm strong})\}$, and it can be observed that turbulence has little effect on $\langle T_{1} \rangle$ and $\langle \sqrt{T_{1}} \rangle$ when the distance is short. However, as the distance increases to a certain level, they begin to decrease, and the stronger the turbulence intensity, the faster $\langle T_{1} \rangle$ and $\langle \sqrt{T_{1}} \rangle$ decrease. The  numerical simulations illustrate that atmospheric turbulence has a significant impact on the transmittance of long-distance CV-QSS.

\begin{table}
\begin{center}
\caption{Default parameters of free-space CV-QSS in simulations.}
\label{table1}
\begin{tabular}{|c|c|c|}
\hline
Symbol& 
Quantity& 
Value 
\\
\hline
 $\lambda_j$& 
Wavelength of $U_{j}$'s Gaussian beam& 
$8\times 10^{-5}$m \\
\hline
$W_{0j}$& 
Initial radius of $U_{j}$'s Gaussian beam& 
0.02 m \\
\hline
$r$& 
Receiving antenna radius& 
0.2 m \\
\hline
$d_{cor}$& 
Diameter of fiber core& 
$9\times 10^{-6}$ m \\
\hline
$D_f$& 
Focal length of collecting lens& 
0.22 m \\
\hline
$L_0$& 
The outer scale of turbulence& 
0.04 m \\
\hline
$f_{pr}$& 
The pulse recurrence frequency & 
$1\times 10^{8}$ Hz\\
\hline
$R_{ra}$& 
The duty ratio& 
0.15 \\
\hline
$\eta$& 
Reconciliation parameter& 
0.95\\
\hline
$\eta_e$& 
The efficiency of the dealer's detector& 
0.5\\
\hline
$T_H$& 
The transmissivity of the HABS& 
0.99\\
\hline
$\epsilon_{0}$& 
Original excess noise introduced by each participant& 
0.01 SNU\\
\hline
$v_{el}$& 
The noise variance of the dealer's detector& 
0.1 SNU\\
\hline
$V_A$& 
$U_{j}$'s modulation variance& 
1 SNU\\
\hline
$h$& 
A height above sea level& 
10 m\\
\hline
$d_{dB,j}$& 
The ratio between the maximal and minimal amplitudes that $U_j$ can output & 
40 dB\\
\hline
$R_{e,j}$& 
The finite extinction ratio of $U_j$'s the amplitude modulator& 
40 dB\\
\hline
$R_{p,j}$& 
The finite extinction ratio of $U_j$'s the polarization beam splitter& 
30 dB\\
\hline
\end{tabular}
\end{center}
\end{table}
\subsection{The total excess noise model}

Similarly, we use $\epsilon\left(\textbf{T}\right)=(\epsilon_1\left(\textbf{T}\right),\cdot\cdot\cdot,\epsilon_j\left(\textbf{T}\right),\cdot\cdot\cdot,\epsilon_n\left(\textbf{T}\right))$ to describe the excess noise of the free-space CV-QSS system. Here $\epsilon_j\left(\textbf{T}\right)$ represents the excess noise of $Link_j$, and it is important to note that this is related to $\textbf{T}$ for the entire CV-QSS, not just $T_j$. This section will examine the various sources of noise that may be encountered, including modulation, photon leakage, phase errors, and transmittance fluctuations.

In the process of preparing the coherent state signal, modulator imperfection causes noise $\epsilon_{am}$. In a CV-QSS system, $n$ participants should prepare coherent states, so the modulation noise consists of $n$ parts. For $Link_j$, this noise referred to the channel input can be modeled as
\begin{equation}\label{am}
\begin{split}
\epsilon_{am,j}(\textbf{T})=\frac{1}{T_j}\sum_{i=1}^n \left(T_i|\alpha_{smax,i}|^210^{-0.1d_{dB,i}}\right),
\end{split}
\end{equation}
where $|\alpha_{smax,i}|^2\approx 10V_{A_i}$ is the maximal amplitude of the $U_i's$ signal pulse, and $d_{dB,i}$ is the ratio between the maximal and minimal amplitudes that $U_j$ can output  \cite{marie2017self}.

There is a photon-leakage noise caused by the leakage from the phase reference pulse to the signal pulse \cite{shao2022phase}. 
For $Link_j$ of CV-QSS, the phase reference of $U_j's$ signal is coupled to all signal pulses from $U_1$ to $U_n$, that is, the $n$ modulated signals may be contaminated by the phase reference of $U_j's$ signal. Therefore, the photon-leakage noise of $Link_j$ in the CV-QSS can be identified as
\begin{equation}\label{le}
\epsilon_{le,j}(\textbf{T})=\frac{2E_{R,j}^2}{T_j}\sum_{i=1}^n \left(T_{i}10^{-0.1(R_{e,i}+R_{p,i})}\right),
\end{equation}
where $E_{R,i}$ is the amplitude of the phase reference $|\alpha^{R}_i\rangle$, $R_{e,i}$ and $R_{p,i}$ are the finite extinction ratios of the amplitude modulator and the polarization beam splitter, respectively.

The LLO form can close the security loopholes brought by the TLO system, but it will introduce nontrivial phase errors to the system. The LO noise of $Link_j$ caused by phase errors is given by \cite{marie2017self}
\begin{equation}\label{5}
\epsilon_{lo}(\textbf{T})=2V_{A_j}(1-e^{-\frac{V_{e,j}}{2}}),
\end{equation}
where $V_{e,j}=V_{p,j}+V_{t,j}+V_{m,j}$ is the variance of the phase noise, which is mainly derived from the phase drift of signal pulse and phase reference in three stages of preparation, transmission  and measurement. We have $V_{p,j}=0$ and $V_{t,j}=0$, when let signal pulse and phase reference be generated from the same optical wave front and transmitted in the same quantum channel \cite{PhysRevA.104.032608}. Therefore, the LO noise mainly comes from phase errors $V_{m,j}$ in the heterodyne detection. In low $V_{m,j}$, the LO noise can be simplified to
\begin{equation}\label{lo}
\epsilon_{lo,j}(\textbf{T})=V_{A_j}V_{m,j}=V_{A_j}\frac{\chi_j(T_j)+1}{E^2_{R,j}},
\end{equation}
where $E_{R,j}$ is the phase-reference amplitude on the dealer's side, and  $\chi_j(T_j)=\frac{1}{T_j}-1+e_0+\frac{2-\eta+2v_{el}}{\eta T_j}$ is the total noise  imposed on the phase-reference with  channel noise of phase reference $e_0$ and electrical noise $v_{el}$. 

Theoretically, when all subchannels are considered, the sum of the above several types of noise of $Link_j$ in the free-space channel is
\begin{equation}\label{oth}
\begin{split}
\langle \epsilon_{oth,j}(\textbf{T})\rangle=\int_{\mathbb{R}^n}\left[\epsilon_{am,j}(\textbf{T})+\epsilon_{le,j}(\textbf{T})+\epsilon_{lo,j}(\textbf{T}) \right]p(\textbf{T})d\textbf{T}.
\end{split}
\end{equation}
According to the Eqs. \ref{am}-\ref{oth} and the fact that the elements in $\textbf{T}$ are independent of each other, the noise $\langle \epsilon_{oth,j}(\textbf{T})\rangle$ actually can be given by
\begin{equation}\label{oth1}
\begin{split}
\langle \epsilon_{oth,j}(\textbf{T})\rangle&=\langle \epsilon_{am,j}(\textbf{T})\rangle+\langle \epsilon_{le,j}(\textbf{T})\rangle+\langle \epsilon_{lo,j}(\textbf{T})\rangle\\
&=\epsilon_{am,j}(\langle\textbf{T}\rangle)+\epsilon_{le,j}(\langle\textbf{T}\rangle)+\epsilon_{lo,j}(\langle\textbf{T}\rangle).
\end{split}
\end{equation}

Furthermore, in a free-space channel, the transmittance exhibits fluctuations over time, resulting in the generation of transmittance fluctuation noise. This noise is associated with the modulation variance of the sender \cite{chai2020suppressing}, as well as the variance of the transmittance, which is indicative of the magnitude of the transmittance fluctuations. Therefore, the fluctuating noise $\epsilon_{tf,j}(\textbf{T})$ can be written as
\begin{equation}\label{5}
\begin{split}
\epsilon_{tf,j}(\textbf{T})&={\rm var}\left(\sqrt{T_{j}}\right)V_{A_j}\\
&=\left(\langle T_{j}\rangle-\langle\sqrt{T_{j}}\rangle^2\right)V_{A_j}\\
&=\langle\epsilon_{tf,j}(\textbf{T})\rangle.
\end{split}
\end{equation}

In conclusion, the expectation of the total excess noise of $Link_j$ in the CV-QSS can be quantified as
\begin{equation}\label{5}
\langle\epsilon_{j}(\textbf{T})\rangle=\langle \epsilon_{oth,j}(\textbf{T})\rangle+\langle\epsilon_{tf,j}(\textbf{T})\rangle+\langle\epsilon_{0,j}(\textbf{T})\rangle,
\end{equation}
where $\epsilon_{0,j}$ is the additional noise from the unidentified or unprotected sources. 

Optimizing noise is an effective way to optimize the performance of continuous-variable quantum cryptosystems. From Eqs. \ref{le} and \ref{lo}, $\epsilon_{le,j}$ is proportional to $E^2_{R,j}$, while $\epsilon_{lo,j}$ inversely is proportional to $E^2_{R,j}$. Therefore, combining with the probabilistic statistical properties of transmittance, $\langle\epsilon_{le,j}\rangle+\langle\epsilon_{lo,j}\rangle$ is theoretically the optimal (minimum) value when 
\begin{equation}\label{5}
E^2_{R,j}=\sqrt{\frac{\langle T_j\rangle V_{A_j}(\chi_j(\langle T_j\rangle)+1)}{2\sum_{i=1}^n \left(\langle T_{i}\rangle 10^{-0.1(R_{e,i}+R_{p,i})}\right)}}.
\end{equation}
\begin{figure}[!t]
\centering
\includegraphics[width=3in]{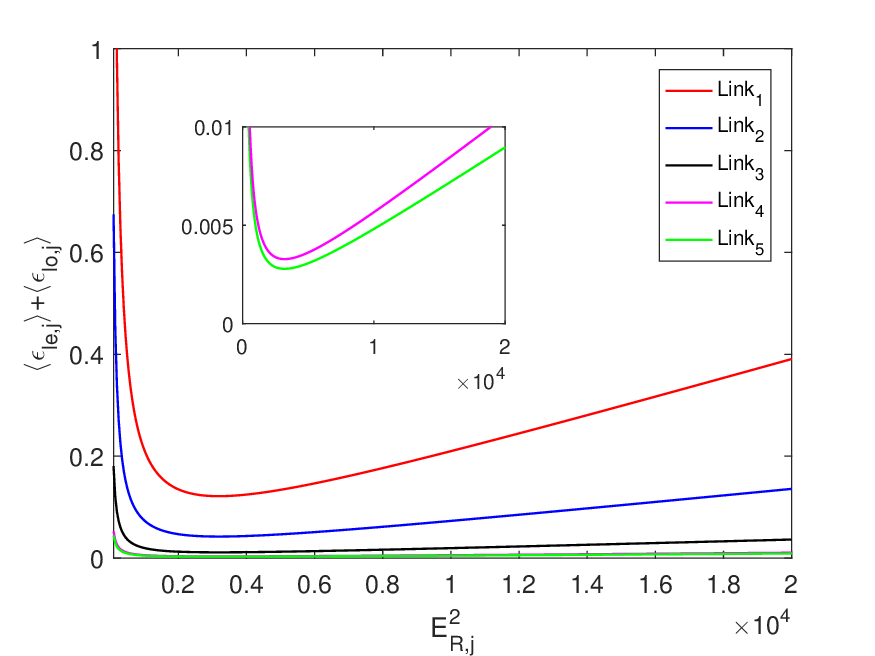}
\caption{The noise $\langle\epsilon_{le,j}\rangle+\langle\epsilon_{lo,j}\rangle$ as a function of $E^2_{R,j}$ with $L=10$ km and $C^2_n=1\times 10^{-15}m^{-2/3}$.}
\label{noise}
\end{figure}
Fig. \ref{noise} visualizes the existence of an optimal $E^2_{R,j}$ for each link of the CV-QSS.

\begin{figure}[!t]
\centering
\includegraphics[width=3in]{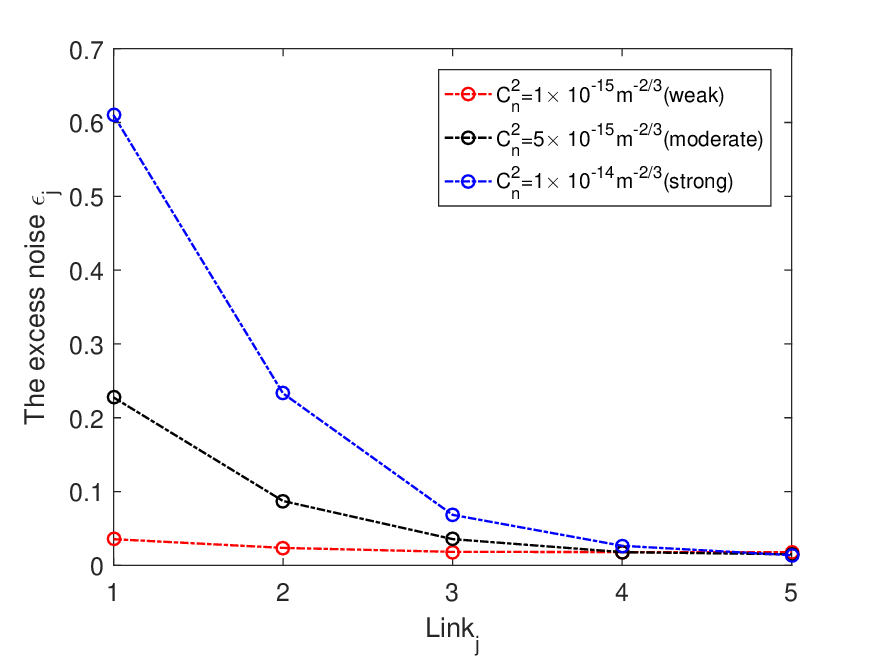}
\caption{The total excess noise $\epsilon_{j}$ of three different turbulence intensities with $n=5$ and $L=10$ km.}
\label{excess}
\end{figure}

From the above modeling of the excess noise, it can be seen that $\langle \epsilon_{oth,j}\rangle$ is inversely proportional to $\langle T_j\rangle$. It is also known from \ref{sec:3.1} that the transmittance $\langle T_1\rangle$ of $Link_1$ is the lowest of $\{\langle T_j\rangle\}$, so $\langle \epsilon_{oth,1}\rangle$ is the largest of $\{\langle \epsilon_{oth,j}\rangle\}$. Therefore, we infer that the total excess noise $\epsilon_j$ of $Link_1$ is the largest among all links in the CV-QSS, assuming $\epsilon_{0,j}=\epsilon_{0}=0.01$. Fig. \ref{excess} shows a numerical simulation of the total excess noise with the optimal $E^2_{R,j}$, which verifies our inference. In addition, for a given link (such as $Link_1$), an increase in turbulence intensity is accompanied by an increase in excess noise.
\subsection{The interruption probability}\label{sectioninterruption}
In a free-space channel, a large angle-of-arrival fluctuation of the signal can, with a certain probability, lead to an interruption of the quantum communication. 
Specifically, the beam  jitters randomly in the receiving lens, where case the focus is also randomly distributed. If the focus lies outside the receiving fiber core, the quantum communication is interrupted.

For CV-QSS, all $n$ QKD links must be non-disruptive in order to generate the $n$ subkeys $\{K_1,\cdot \cdot \cdot,K_n\}$ in the key $K$, thus the non-interruption probability of the whole CV-QSS system is $P^{non}_{QSS}=\prod \limits_{j=1}^n(1-P_j)$, where $P_j$ is the interruption probability of $Link_j$.
Consequently, the interruption probability of the CV-QSS system can be obtained as
\begin{equation}
P_{QSS}=1-P^{non}_{QSS}=1-\prod \limits_{j=1}^n(1-P_j).
\end{equation}
For the QKD between the participant $U_j$ and the dealer ($Link_j$), the interruption probability is related to the angle-of-arrival $\theta_{aj}$, fiber core $d_{core}$, and transmission distance $L_j$. The interruption probability of $Link_j$ can be expressed as \cite{wang2018atmospheric}
\begin{equation}
\small
P_j=1-\int_{\frac{{\rm -d_{core}}}{2}}^{\frac{{\rm d_{core}}}{2}}\frac{1}{D_f\sqrt{2\pi \langle\theta_{aj}^2\rangle}}{\rm exp}\left[\frac{-x^2}{2D_f^2\langle\theta_{aj}^2\rangle}\right]{\rm d}x,
\end{equation}
where $D_f$ is the focal length and the variance of $\theta_{aj}$ is
\begin{equation}
\langle\theta_{aj}^2\rangle=\frac{\langle x_{0j}^2\rangle}{L_j^2}.
\end{equation}

\begin{figure}[!t]
\centering
\includegraphics[width=3in]{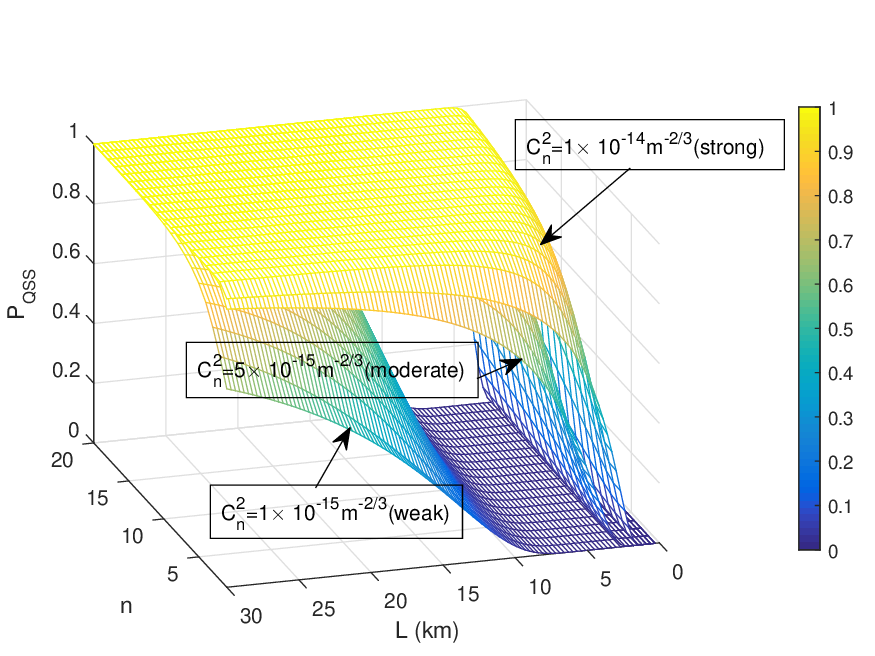}
\caption{The interruption probability as a function of distance L (km) and number of participants $n$ in different atmospheric turbulence intensities: $C^2_n=1\times 10^{-14}m^{-2/3}$(strong), $C^2_n=5\times 10^{-15}m^{-2/3}$(moderate), and $C^2_n=1\times 10^{-15}m^{-2/3}$(weak).}
\label{interruptionf}
\end{figure}

Fig. \ref{interruptionf} shows a three-dimensional diagram of the interruption probability as a function of distance and number of participants. For a fixed turbulence intensity, the probability of interruption is observed to increase in direct correlation with the number of participants involved and the transmission distance. At weak turbulence intensity ( $C^2_n=1\times 10^{-15}m^{-2/3}$), the interruption probability grows slowly, while when turbulence intensity is strong ($C^2_n=1\times 10^{-14}m^{-2/3}$), the interruption probability quickly approaches the value of 1. This fact shows that as the intensity of turbulence rises, the interruption probability increases, even causing the system to be completely disrupted at short distances and at small scales (the number of participants).


\section{The secret key rate}\label{sec:4}
In this section, we present the calculation method of the secret key rate for free-space CV-QSS and offer a numerical simulation-based assessment of its performance.
Section \ref{sectioninterruption} indicates that the communication interruption probability due to angle of arrival fluctuations cannot be disregarded. Moreover, as mentioned in Section \ref{sec:2}, CV-QSS can be decomposed into $n$ independent QKD links. In order to ensure the security of the whole non-interruptible CV-QSS system, the minimum value in $\{r_1,\cdot \cdot \cdot\,r_n\}$ should be selected as the lower bound of the system key rate. Therefore, the lower bound of secret key rate for the free-space CV-QSS can be given as
\begin{equation}\label{rqss}
\begin{split}
R&=\left(1-P_{QSS}\right)\times{\rm min}\{r_1,\cdot \cdot \cdot\ ,r_n\}\\
&=\prod \limits_{j=1}^n(1-P_j)\times{\rm min}\{r_1,\cdot \cdot \cdot\ ,r_n\}.
\end{split}
\end{equation}


According to the security analysis theory of GMCS CV-QKD \cite{2012Gaussianquantuminformation}, the secret key rate of $Link_j$ is given as
\begin{equation}
r_j=\eta I_{U_jD}-\chi_{ED},
\end{equation}
where $I_{U_jD}$ is the Shannon mutual information between $U_j$ and the dealer, and $\chi_{ED}$ is the Holevo bound, which is the maximum information that Eve can obtain based on the dealer's variable. 
From the knowledge of information theory, the Shannon mutual information can be calculated by
\begin{equation}
I_{U_jD}=\log \frac{V+\chi^t_{j}}{1+\chi^t_{j}},
\end{equation}
where the overall noise referred to the channel input is given by \cite{grice2019quantum}
\begin{equation}\label{5}
\chi^t_{j}=\chi^l_{j}+\chi_h/T^e_{j}.
\end{equation}
Here 
\begin{equation}
\chi^l_{j}=\frac{1}{T^e_{j}}-1+\langle\epsilon_{j}\rangle
\end{equation}
is the channel-added noise and $\chi_h=\frac{2-\eta_e+2v_{el}}{\eta_e}$ is the noise caused by the dealer's heterodyne detection, where $\eta_e$ and $v_{el}$ are the detection efficiency and the electronics noise of the detection, respectively. Note that $T^e_{j}$ represents the equivalent transmittance of $Link_j$ in a free-space channel, where $T^e_{j}=\left\langle\sqrt{T_j}\right\rangle^2$ \cite{hosseinidehaj2021composable}. There is already a generally accepted model \cite{PhysRevA.76.042305} for how to compute the Holevo bound, which can be written as
\begin{equation}\label{chi}
\chi_{ED}=\sum_{i=1}^2G(\nu_i)-\sum_{i=3}^5G(\nu_i),
\end{equation}
where $G(\nu)=\frac{\nu+1}{2}\log_2\frac{\nu+1}{2}-\frac{\nu-1}{2}\log_2\frac{\nu-1}{2}$.
The method for calculating symplectic eigenvalues can be referred to in Appendix B of \cite{yang2023continuous}.

For the CV-QSS system, the secret key rate is calculated by Eq. \ref{rqss}, where $P_{QSS}$ was discussed in Section \ref{sectioninterruption}. The next point is to discover ${\rm min}\{r_1,\cdot \cdot \cdot\,r_n\}$. \cite{yang2023continuous} has presented a comparison among $n$ key rates in a uniform fast-fluctuating channel and proved that $r_1$ is the smallest one of them. Actually, for the $n$ QKD links formed by the dealer with $U_1, \cdot \cdot \cdot, U_j$, respectively, we assume that the values of the same parameters (such as the modulated variance) are the same for all links except for the equivalent transmittance and the total excess noise, which are the two most relevant parameters for the secret key rate, according to an in-depth study of two-party CV-QKD \cite{2012Gaussianquantuminformation,Denys2021explicitasymptotic}. Finding the QKD link with the lowest key rate is actually finding the link with the lowest transmittance and largest excess noise \cite{liao2023continuous}. According to the previous analysis, $Link_1$ satisfies the two requirements. Therefore, the key rate of CV-QSS can be described as 
\begin{equation}
R= \left[1-P_{QSS}\right]\times r_1.
\end{equation}

\begin{figure}[!t]
    \centering
    \subfloat[\small Secret key rates as the function of distance L with $n=5$.]{\includegraphics[width=3in]{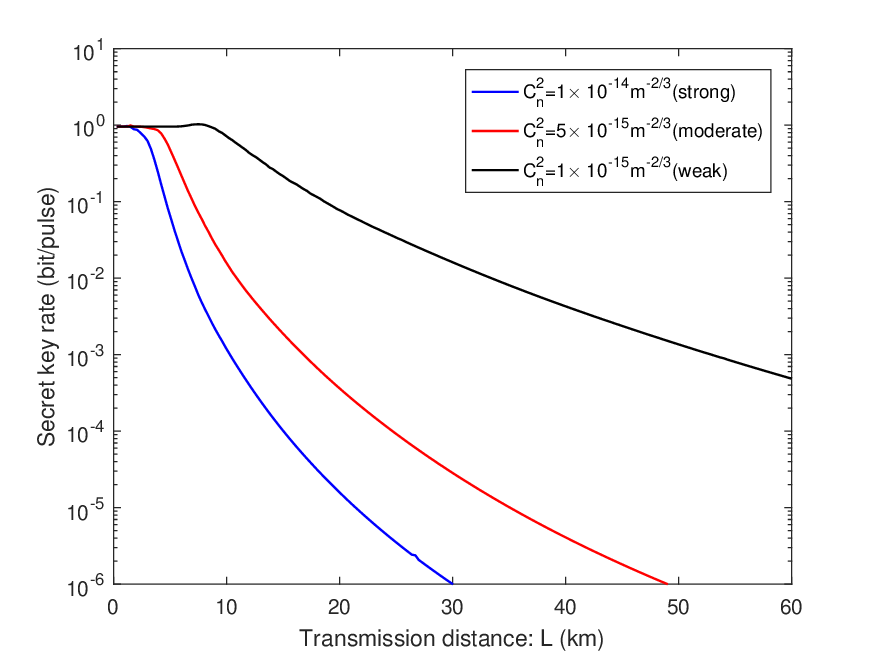}}\\%
    \label{qkdlink1}
    \subfloat[\small Key rates as the function of the total number of participants n with $L=10$ km.]{\includegraphics[width=3in]{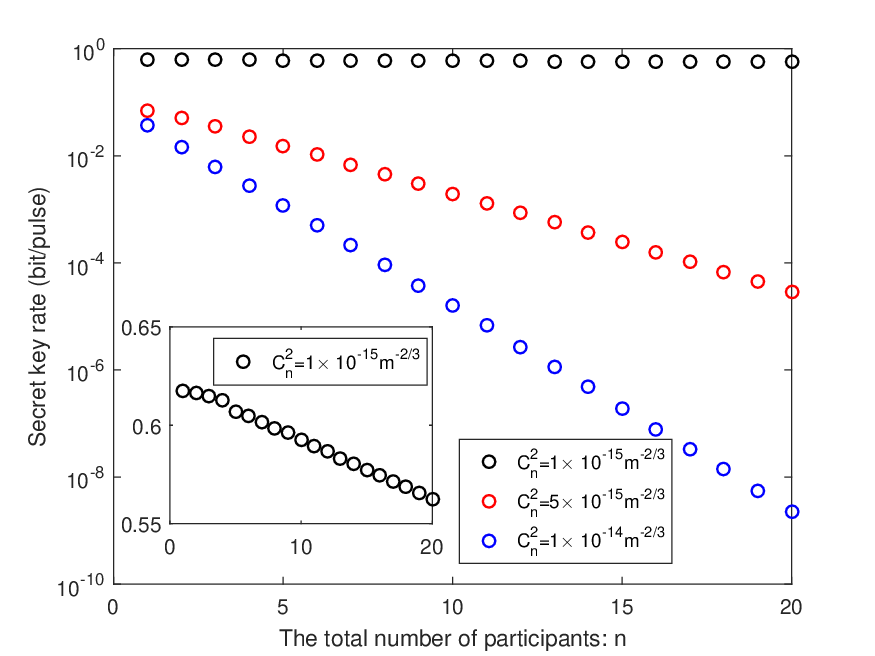}}%
    \label{qkdlink2}
    \caption{The secret key rate of free-space CV-QSS with different atmospheric turbulence intensities: $C^2_n=1\times 10^{-14}m^{-2/3}$(strong), $C^2_n=5\times 10^{-15}m^{-2/3}$(moderate), $C^2_n=1\times 10^{-15}m^{-2/3}$(weak).}
\label{keyratefzt}
\end{figure}


We analyze the performance of CV-QSS through numerical simulations. As shown in Fig. \ref{keyratefzt}(a) with the number of participants $n = 5$, the key rate exhibits a decline in conjunction with an augmentation in transmission distance, when the same turbulence intensity ($C^2_n$) is maintained. As the turbulence intensity rises, the key rate declines at a more pronounced rate with distance. This is due to the fact that as the turbulence intensity increases, the equivalent transmittance declines, the probability of interruption rises, and the level of channel noise rises. Nevertheless, it is evident that at a turbulence intensity of $C^2_n=1\times 10^{-15}m^{-2/3}$, the key rate can still reach approximately $5\times 10^{-3}$ bit/pulse when the transmission distance reaches $60$ km, indicating that the maximum transmission distance is greater than $60$ km.
As can be seen from Fig. \ref{keyratefzt}(b), for increasing $n$, the key rate of the CV-QSS system decreases, especially when the turbulence intensity is strong ($C^2_n=1\times 10^{-14}m^{-2/3}$). The underlying reason for this is that an increase in the number of participants directly leads to an increase in the total channel noise and the system's interruption probability. 
Nevertheless, when turbulence intensity reaches the level of $C^2_n=1\times 10^{-15}m^{-2/3}$, the impact of the number of participants on the key rate is not readily discernible. Even when the number of participants reaches a value of $n = 20$, the key rate can still remain above 0.55 bit/pulse when the transmission distance is $L=10$ km.

\section{Conclusions}\label{sec:5}

In this paper, we have introduced a free-space CV-QSS scheme based on LLO and conducted an analysis of its characteristics with respect to channel transmittance, excess noise, and interruption probability. These factors are essential for determining the practical secret key rate of the CV-QSS system.

Initially, two primary aspects of free-space beam extinction and atmospheric turbulence were considered, and an elliptic model and Monte Carlo method were employed to obtain the correlation expectation of CV-QSS transmittance $\textbf{T}$.
Particularly, we conducted an evaluation of the various sources of excess noise based on the structural attributes of CV-QSS with LLO. These sources include imperfect signal modulation, photon leakage, non-trivial phase errors resulting from LLO, and transmittance fluctuation. Moreover, the optimal $E^2_{R}$ was demonstrated and provided to optimize the excess noise.
Then, we derived a method for calculating the interruption probability of the CV-QSS system. On the basis of the above work, the key rate bound of the free-space CV-QSS was determined.

In the analysis of system performance, it was observed that an increase in turbulence intensity and the number of participants results in a decrease in both the key rate and transmission distance. This is primarily due to the fact that as $C_n^2$ and $n$ increase, the equivalent transmittances of the system decrease while the probability of interruption and total channel noise increase. 
When the turbulence intensity is $C_n^2=1\times 10^{-15}m^{-2/3}$, our protocol is shown to have the capacity to securely share secrets over a distance of more than $60$ km between five participants, while still maintaining a high key rate of $0.55$ bit/pulse over a distance of $10$ km across twenty participants.
These findings indicate that the realization of efficient quantum secret sharing in free space is viable. Future research will focus on strategies to enhance channel the transmittance expectation and reduce the probability of interruption and channel noise, thereby improving overall system performance.

\begin{appendices}
\section{The parameters of $T_{at,j}$}
We show some details on the elliptic-beam model for $T_{at,j}$. The maximal transmittance for a centered beam can be given by
\begin{equation}\label{}
\begin{split}
T_{0j}&=1-I_0\left(r^2\left[W_{1j}^{-2}-W_2^{-2}\right]\right){\rm exp}^{-r^2\left(W_{1j}^{-2}+W_{2j}^{-2}\right)}\\
&-2\left\{1-{\rm exp}\left[-\frac{r^2}{2}\left(W_{1j}^{-1}-W_{2j}^{-1}\right)^2\right]\right\}
{\rm exp}\left\{-\left[\frac{\frac{(W_{1j}+W_{2j})^2}{W_{1j}^2-W_{2j}^2}}{R(W_{1j}^{-1}-W_{2j}^{-1})}\right]^{Q(W_{1j}^{-1}-W_{2j}^{-1})}\right\}
\end{split}
\end{equation}
with the modified Bessel function of i-th order $I_i(\cdot)$, where $R(\cdot)$ and  $Q(\cdot)$ are scale and shape functions, respectively,
\begin{equation}
R(x)=\left[{\rm ln}\left(2\frac{1-{\rm exp}(-r^2x^2/2)}{1-{\rm exp}(-r^2x^2)I_0(r^2x^2)}\right)\right]^{-1/Q(x)},
\end{equation}
\begin{equation}
\begin{split}
Q(x)&=2r^2x^2\frac{{\rm exp}(-r^2x^2)I_1(r^2x^2)}{1-{\rm exp}(-r^2x^2)I_0(r^2x^2)}
\left[{\rm ln}\left(2\frac{1-{\rm exp}(-r^2x^2/2)}{1-{\rm exp}(-r^2x^2)I_0(r^2x^2)}\right)\right]^{-1}.
\end{split}
\end{equation}
${\rm W_{eff}}(\cdot)$ is the effective squared spot radius written as
\begin{equation}
{\rm W_{eff}}(x)=2r\left[\textbf{\emph{W}}\left(f_1(x)\frac{4r^2}{W_{1j}W_{2j}}f_2(x)\right)\right]^{-\frac{1}{2}},
\end{equation}
where $f_1(x)={\rm exp}[(r^2/W_{1j}^2)(1+2\cos^2x)]$, $f_2(x)={\rm exp}[(r^2/W_{2j}^2)(1+2\sin^2x)]$, and $\textbf{\emph{W}}(\cdot)$ is the Lambert {\emph{W}} function \cite{Corless1996}.

\section{The covariance matrix elements of $\textbf{w}$}\label{elementsw}
Gaussian independent variables $x_{0j}$ and $y_{0j}$ have no correlations with Gaussian variables $\phi_{1j}$ and $\phi_{2j}$. However, there is a correlation between the latter two variables. The mean values $\langle x_{0j}\rangle$ and $\langle y_{0j}\rangle$ are assumed to be zero, then $\textbf{w}$ can be described by a covariance matrix

\begin{equation}
\gamma_w=\left(\begin{array}{cccc} 
    \langle x_{0j}^2\rangle &    0    & 0&0 \\ 
    0 &    \langle y_{0j}^2\rangle   & 0&0\\ 
    0 & 0 & \langle \phi_{1j}^2 \rangle &\langle \phi_{1j}\phi_{2j} \rangle\\
0 & 0 & \langle \phi_{1j}\phi_{2j}\rangle & \langle \phi_{2j}^2 \rangle
\end{array}\right), 
\end{equation}
where the diagonal elements of the covariance matrix associated with $x_{0j}$ and $y_{0j}$ are given by\cite{PhysRevA.96.043856}
\begin{equation}
\langle x_{0j}^2\rangle=\langle y_{0j}^2\rangle=0.33W_{0j}^2\sigma_{lj}^2\Omega_j^{-6/7}.
\end{equation}
$\Omega_j=k_jW^2_{0j}/2L_j$ is the Fresnel parameter and 
\begin{equation}
\sigma_{lj}=1.23C_n^2k_j^{7/6}L_{j}^{11/6}
\end{equation}
is the Rytov variance. 
Here $C_n^2$ is the index of refraction structure parameter, and it describes the strength of turbulence. $k_j = 2\pi/\lambda_j$ is the optical wave number of light with wavelength $\lambda_j$. 
The other covariance matrix elements of $\textbf{w}$ related to variables $\phi_{1j}$ and $\phi_{2j}$  are described as
\begin{equation}
\langle \phi_{1j,2j} \rangle={\rm ln}\frac{(1+2.96\sigma_{lj}^2\Omega_j^{5/6})^2}{\Omega_j^{2}\sqrt{(1+2.96\sigma_{lj}^2\Omega^{5/6})^2+1.2\sigma_{lj}^2\Omega_j^{5/6}}},
\end{equation}
\begin{equation}
\langle \phi_{1j,2j}^2 \rangle={\rm ln}\left(1+\frac{1.2\sigma_{lj}^2\Omega_j^{5/6}}{(1+2.96\sigma_{lj}^2\Omega_j^{5/6})^2}\right),
\end{equation}
\begin{equation}
\langle \phi_{1j}\phi_{2j}\rangle={\rm ln}\left(1-\frac{0.8\sigma_{lj}^2\Omega_j^{5/6}}{(1+2.96\sigma_{lj}^2\Omega_j^{5/6})^2}\right).
\end{equation}
\nocite{*}
\end{appendices}

\newpage


\vfill

\end{document}